\documentclass{aa}  
\usepackage{natbib}
\usepackage{graphicx}
\usepackage{multirow}
\usepackage[table]{xcolor}
\definecolor{lightgray}{gray}{0.9}
\usepackage{txfonts}
\begin{document}

\title{Dust as interstellar catalyst - II.}
\subtitle{How chemical desorption impacts the gas}
\author{S. Cazaux\inst{1,2}, M. Minissale\inst{3}, F. Dulieu\inst{3},  S. Hocuk\inst{4}} 
\offprints{cazaux@astro.rug.nl}
\institute{Kapteyn Astronomical Institute, University of Groningen, P.O. Box 800, 9700AV Groningen, The Netherlands
\and
Leiden Observatory, Leiden University, P.O. Box 9513, NL 2300 RA Leiden, The Netherlands
\and
LERMA, UMR 8112 du CNRS, Observatoire de Paris et Universit\'e de Cergy Pontoise, 5 Mail Gay-Lussac, F-95031 Cergy-Pontoise Cedex, France
\and
Max Planck Institute for Extraterrestrial Physics, Giessenbachstr. 1, D-85741, Garching, Germany}

\date{Received ; accepted }

%%%%%%%%%%%%%%%%%%%%%%%%%%%%%%%%%%%%%%%%%%%%%%%%%%%%%%%%%%%%%%%%%
\abstract {Interstellar dust particles, which represent 1$\%$ of the total mass, are recognized to be very powerful interstellar catalysts in star-forming regions. The presence of dust can have a strong impact on the chemical composition of molecular clouds. While observations show that many species that formed onto dust grains populate the gas phase, the process that transforms solid state into gas phase remains unclear.}{The aim of this paper is to consider the chemical desorption process, i.e. the process that releases solid species into the gas phase, in astrochemical models. These models allow determining the chemical composition of star-forming environments with an accurate treatment of the solid-phase chemistry.}{In paper I we derived a formula based on experimental studies with which we quantified the  efficiencies of the chemical desorption process. Here we extend these results to astrophysical conditions.}{The simulations of astrophysical environments show that the abundances of gas-phase methanol and H$_2$O$_2$  increase by four orders of magnitude, whereas gas-phase H$_2$CO and HO$_2$  increase by one order of magnitude when the chemical desorption process is taken into account. The composition of the ices strongly varies when the chemical desorption is considered or neglected. }{We show that the chemical desorption process, which directly transforms solid species into gas-phase species, is very efficient for many reactions. Applied to astrophysical environments such as $\rho$ Oph A, we show that the chemical desorption efficiencies derived in this study reproduce the abundances of observed gas-phase methanol, HO$_2$, and H$_2$O$_2$, and that the presence of these molecules in the gas shows the last signs of the evolution of a cloud before the frost. }


\titlerunning{Dust as interstellar catalyst II}
\maketitle

\section{Introduction}

Observations of cold environments (i.e., pre-stellar cores) have shown unexpected amounts of several molecules in the gas phase (\citealt{Bacmann2012}) while these species should be depleted in gas and adsorbed onto cold dust grains (T$<$20 K) (\citealt{collings2004}). To understand these observations, many experimental studies have focused on the transition between solid and gas phase in conditions similar to the one encounter in the interstellar medium (ISM). Many mechanisms involving thermal processes (\citealt{collings2004}, \citealt{bisschop2006}) or non-thermal processes, such as photodesorption (\citealt{DeSimone2013}; \citealt{bertin2013}) and sputtering (\citealt{Johnson2013}; \citealt{cassidy2013}) have been studied. Recently, \cite{dulieu2013} showed that species forming onto dust particles could be directly ejected into the gas phase upon formation. This mechanism, called chemical desorption, links solid and gas phases without the intervention of external agents such as photons, electrons, or other energetic particles. This process could therefore be efficient in environments where UV and cosmic rays (CR) are shielded, and where photodesorption or sputtering cannot account for the release of solid species into the gas phase. 

Observations of $\rho$ Oph A reported the presence of species such as O$_2$, HO$_2$, and H$_2$O$_2$ (\citealt{bergman2011}; \citealt{parise2012}; \citealt{liseau2012}) that could be attributed to the effect of warm dust allowing the evaporation of their ices. In particular, the detection of H$_2$O$_2$ in $\rho$ Oph A, a molecule that is known to be present in ices but has never
been observed before, raised the question of its origin. \cite{du2012} modeled the abundances of gas and ice species in environments with similar physical characteristics as $\rho$ Oph A. These authors could reproduce the abundance of H$_2$O$_2$ by assuming a chemical desorption efficiency of 7$\%$, derived from the approach of \cite{garrod2007}. 

In \cite{minissale2015b} (hereafter Part I of this study), we performed several experiments to measure and quantify the chemical desorption efficiency. We found that this process is mainly sensitive to four parameters: enthalpy of formation, degrees of freedom, binding energy, and mass of newly formed molecules. We quantified the chemical desorption process as a function of these parameters, and we here determine its effect on the chemical composition of an evolving cloud. We re-evaluate the abundances of gas-phase O$_2$, HO$_2$, and H$_2$O$_2$ with the chemical desorption efficiencies derived in this study and consider the change in these efficiencies when the dust surfaces are covered by ice.

%Our study is motivated by the recent detections of O$_2$, HO$_2$ and H$_2$O$_2$ in $\rho$ oph A (\citealt{bergman2011}; \citealt{parise2012}; \citealt{liseau2012}) and of O$_2$ in Orion (\citealt{goldsmith2011}). The detection of O$_2$ in these two disctinct environemnts have been atributted to the effect of warm dust grains that allow the evaporation of the ices covering the dust. Additionally, the recent detection of H$_2$O$_2$ in $\rho$ oph A, a molecule present in ices but never observed before, raised the question of its origin. \cite{du2012} modelled the abundances of gas and icy species in environments with similar physical characteristics than $\rho$ oph A. These authors reproduce the abundance of H$_2$O$_2$ by assuming a chemical desorption efficiency of 7$\%$, derived from the approach of \cite{garrod2007}. Here we re-evaluate the abundances of O$_2$, HO$_2$ and H$_2$O$_2$ with the chemical desorption efficiencies derived in this study, and the change in these efficiencies when the dust surfaces are covered by ice.

This paper is organized as follows: in Sect.~2 we present the law that has been determined to quantify the chemical desorption process, as shown in part I. In Sect.~3 we describe the chemical model that is used to simulate the chemical composition of a cloud when chemical desorption is considered. In Sect.~4 we discuss the necessary careful treatment of the solid phase, which takes into account the chemical desorption process to reproduce the observations of specific chemical species in $\rho$ Oph A. In Sect. 5 we present the main conclusions of this study.

%Oxygen is the most abundant “metal” element in the cosmos (Savage \& Sembach 1996; Asplund et al. 2009). In the cold dense interstellar clouds, gas-phase chemical models predict that oxygen mainly resides in CO and O2 molecules (Herbst \& Leung 1989; Millar \& Herbst 1990; Wakelam et al. 2006). However, although CO is ubiquitously distributed in the inter- stellar medium, O2 is not. The latter is only detected very re- cently in ρ Oph A at a low abundance (relative to molecular hydrogen) of 5 × 10−8 (Larsson et al. 2007), and in Orion at an abundance of (0.3–7) × 10−6 (Goldsmith et al. 2011). On the other hand, the observed water (gas or ice) abundance can be as high as 10−4 (van Dishoeck 2004). Thus it seems that water, instead of O2, is a main reservoir of oxygen in addition to CO. When only gas phase chemistry is included, the H2O can be on the order 10−7 at most (see, for example, Bergin et al. 2000; Roberts \& Herbst 2002) for typical dark cloud conditions. That O2 is overproduced and H2O is underproduced in gas phase chemistry suggests that adsorption onto the grain surfaces and the reactions on the surfaces may play important roles.

\begin{table*}
\caption{Reactions adopted in our model with chemical desorption efficiencies for reactions on bare and icy surfaces.}
\label{reac}
\begin{tabular}{cc||c||cc||c||c}
Reaction & Desorbing & Delta H$_R$ (eV) & \multicolumn{2}{c}{Experimental} & Theoretical&Reaction barrier (K)\\
&  & & CD$_{ice}$(exp){\tiny$^{*}$} &CD$_{bare}$(exp){\tiny$^{*}$} & CD$_{bare}$(th)&\\ \hline
O+H      & OH  &4.44&   25$\pm$15    & 50$\pm$25 & 39&0\\
OH+H     & H$_2$O  &5.17&     30$\pm$15    & 50$\pm$25  & 27&0\\ 
O$_2$+H  & O$_2$H      &   2.24& $<$8   & 10$\pm$10&1.4&0$^{a,b}$\\
O$_2$H+H & H$_2$O$_2$  & 3.69&   $<$8   & $<$5     &0.5&0\\
O$_2$H+H & 2 OH        &  1.47& $<$8    &$<$5   &0.3&0\\
H$_2$O$_2$+H& H$_2$O+OH& 2.95& $<$5    & $<$5 &2.1&1000$^e$\\
O$_3$+H    & O$_2$+OH  &3.33&  $<$10 & $<$8 &8&480$^{c,d}$\\
O+O        & O$_2$     & 5.16&$<$5   & 40-80&68&0\\
O$_2$+O    & O$_3$     & 1.1&$<$5   & $<$5 & 0&0\\
N+N    & N$_2$  & 9.79&$>$50  & $>$70 & 89&0\\
CO+H & HCO &0.66& --   & 10$\pm$8& 0.7&2000$^{f}$\\
HCO+H& CO+H$_2$  &3.85& -- & 40$\pm$20 & 47&130\\
HCO+H& H$_2$CO   & 3.91& -- & $<$8 &7&0\\
H$_2$CO+H & CH$_3$O  & 0.88& -- & $<$8&0&2000$^f$\\
H$_2$CO+H& HCO+H$_2$  & 0.61& -- & 10$\pm$5& 0&2200\\
CH$_3$O+H& CH$_3$OH   &4.56& $<$8   & $<$8 &2.3&0\\
CO+O& CO$_2$ & 5.51& $<$5   &  $<$5&22&650$^i$\\
H$_2$CO+O& CO$_2$+H$_2$ &5.45& $<$10   & $<$10&8&350$^i$\\
CH$_3$O+H& H$_2$CO + H$_2$  & 3.64&-- & -- &1.3&150\\
CH$_3$OH + H& CH$_3$O+H$_2$&-0.04&-- & --&0&3200\\ 
OH+O& O$_2$ + H &0.72&-- & --&1.9&0\\
HO$_2$+O&O$_2$+OH&2.2&-- & --&17&0\\
O$_3$+O& O$_2$+O$_2$&4.06&-- & --&39&2500$^g$\\
O+H$_2$v=1& OH+H &-0.08& -- & -- &0&4600$^h$\\
CO$_2$+H&CO+OH&-1.0& --&--&0&10000\\
O+HCO&CO$_2$+H&4.85&--&--&10&0$^i$\\
O+H$_2$CO& CO$_2$+H$_2$&5.45&--&--&8&350$^j$\\
OH+H$_2$& H$_2$O+H &0.65&--&--&0&2100$^k$\\
OH+OH&H$_2$O$_2$&2.21&--&--&0&0\\
OH+CO& CO$_2$+H&1.07&--&--&0&400$^l$\\
OH+HCO&CO$_2$+H$_2$&4.93&--&--&6&0 \\
HO$_2$+H$_2$&H$_2$O$_2$+H&-0.82&--&--&0&5000$^e$\\
H$_2$O+H&OH+H$_2$&-0.65&--&--&0&9600\\
H$_2$O$_2$+H&H$_2$O+OH&2.95&--&--&2&1000\\
\hline\hline
\end{tabular}
\\
\rm{Note that the table is longer than in part I. Here we report the theoretical CD efficiency for more reactions than have been measured experimentally.}\rm
$^a$ \citealt{mokrane2009}, $^b$ \citealt{chaabouni2012}, $^c$ \citealt{mokrane2009}, $^d$ \cite{romanzin2011}, $^e$ \citealt{lamberts2013}, $^f$ \citealt{awad2005}, $^g$ \cite{minissale2014},$^h$ \citealt{weck2005}, $^i$ \cite{minissale2013} from ER, $^j$ \cite{minissale2015}, $^k$ \citealt{oba2012}, $^l$ \citealt{noble2011} 
%\multicolumn{4}{c}{$^a$ Asvany et al. 2008, $^b$ McCall et al. 2004}
$^{*}$ work reported in part I.  
\end{table*}

\section{Chemical desorption process}
In part I of this study, experiments have been performed using the FORMOLISM setup (\citealt{amiaud2006}; \citealt{congiu2009}) to quantify the chemical desorption process. A typical experiment proceeds as follows: one layer (or less) of molecules is deposited at low temperatures ($\sim$ 10~K) on amorphous silicates, oxidized graphite, or compact amorphous water substrate. Immediately after deposition, another beam is aimed at the solid sample and the desorption flux is monitored. These measurements, referred to as \rm{during-exposure desorption (DED),}\rm\ allow determining the efficiency of the chemical desorption process for many of the considered reactions. In a second phase, called temperature-programmed desorption experiments (TPD), the surface temperature is increased and the desorption flux is monitored, which allows determining the amount of formed products that did not desorb upon formation. By combining TPD and DED experiments for selected reactions on several surfaces, we were able to obtain a precise measurement of the chemical desorption efficiency for each reaction. The different reactions are reported in part I of this study.
In Table~\ref{reac} we summarize the different reactions that have been studied with their enthalpy of formation and the experimental chemical desorption efficiency that has been derived on bare (amorphous silicates, oxidized graphite) and icy surfaces.

In part I, we used a simple assumption of equipartition of energy to reproduce the chemical desorption observed experimentally, assuming that the total energy budget $\Delta H_R$ (see Table~\ref{reac}) is shared between all the degrees of freedom $N$. To desorb from the surface, the newly formed products bound with binding energies $E_{bind}$ have to gain velocity in the direction perpendicular to the surface. By considering an elastic collision, the fraction of kinetic energy retained by the product $m$ colliding with the surface, which has an effective mass $M,$ can be determined as
\begin{equation}
\epsilon=\frac{(M-m)^2}{(M+m)^2},
\end{equation}
where we considered an effective mass $M$ of 120  a.m.u. Therefore the total chemical energy available for the kinetic energy perpendicular to the surface is  $\epsilon \Delta H_R/N$. \rm{We assumed that the only velocity perpendicular to the surface corresponds to a distribution of speed or temperature such as kT = $\epsilon \Delta H_R/N$.}\rm\ The probability of the product to have an energy (or liberation velocity) higher than the binding energy becomes

\begin{equation}
CD_{bare}(th) = e^{-\frac{E_{binding}}{\epsilon \Delta H_R /N}}
.\end{equation}

For a water substrate (np-ASW), the experimental results show that the chemical desorption becomes much less efficient and that most of the reactions cannot be determined. Only the reactions O+H, OH+H, and N+N show efficiencies of 25$\%$, 30$\%$, and 50$\%$, respectively.

\section{Chemical model}
We used a three-phase chemical model that combines gas-phase chemistry with surface and bulk chemistry. The grain surface chemistry model (surface + bulk) takes into account the different binding energies of the species on bare or icy surfaces and includes evaporation, photodissociation, and photodesorption processes, which transform surface species either into other surface species or into gas-phase species. \rm{Our model does not take into account the diffusion from species from bulk to surface and from surface to bulk, however. In this sense, when the coverage has reached one layer, the accreting species become bulk species with higher energies and lower diffusion rates (because the diffusion depends on the binding energy).}\rm\ The gas-phase chemical model was adopted from the KIDA database (\citealt{wakelam2012}).

\subsection{Grain surface chemistry}
Our grain surface chemistry model considers the following species: H, H$_2$, O, O$_2$, O$_3$, OH, HO$_2$, H$_2$O, H$_2$O$_2$, CO, CO$_2$, HCO, H$_2$CO, CH$_3$O, and CH$_3$OH, \rm{adapted from \cite{hocuk2015}}\rm. Their binding energies on bare and icy surfaces are listed in Table~\ref{bind}. In the submonolayer regime (coverage $\sim$ 0$\%$) the surface is considered bare, and the binding energies correspond to E$_{bare}$. As the surface becomes covered with ices, we estimate the fraction of the species bound to the bare surface, with  E$_{bare}$, and the fraction bound to the ice with E$_{ice}$. In the multilayer regime (coverage $\ge$ 100$\%$), the surface is considered to be covered by ices, and the binding energies correspond to E$_{ice}$. Our chemical network combines surface and gas-phase species. Gas-phase species are calculated in units of cm$^{-3}$, while surface or bulk species are usually calculated in monolayers (1 ML = 100$\%$ surface coverage). To convert cm$^{-3}$ into monolayers, division by n$_{\rm{dust}}$n$_{sites}$ is needed, which is the number of sites on the dust surface per cm$^{-3}$. This can be written as
\begin{equation}
n_{\rm{dust}}\times n_{sites}=n_{\rm{dust}}\times \frac{4 \pi r^2}{a_{ss}^2}= 4.4\times 10^{15} \times n_{\rm{dust}}\times \sigma,
\end{equation}
where a$_{ss}$ is the distance between two sites, typically 3 \AA, and n$_{\rm{dust}}\times \sigma$ is the total cross section of dust, which is $\sim$ 10$^{-21}$ n$_{\rm{H}}$ cm$^{-2}$ following \cite{mathis1977} (hereafter MRN), and therefore n$_{\rm{dust}}$$\times$ n$_{sites}\sim$ 4.4 $\times$ 10$^{-6}$ n$_{\rm{H}}$. We here report surface species and reactions rates in cm$^{-3}$ to allow a direct comparison with the gas phase. The density of the surface species shown in our calculations can therefore be converted into monolayers by dividing by n$_{\rm{dust}}$$\times$ n$_{sites}$.

\begin{table} [ht]
\caption{Binding energies on bare and icy surfaces. \label{bind}}
\begin{tabular}{lll}
Species & E$_{bare}$ &E$_{ice}$\\
\hline\hline
H&550$^{a,b}$&650$^c$\\
H$_2$&300$^{a,d}$&500$^e$\\
O&1500$^b$&1400$^f$\\
OH&4600$^g$&4600$^g$\\
H$_2$O&4800$^g$&4800$^h$\\
O$_2$&1250$^{i}$&1200$^i$\\
O$_3$&2100$^j$&2100$^j$\\
HO$_2$&4000$^g$&4000$^g$\\
H$_2$O$_2$&6000$^g$&6000$^g$\\
CO&1100$^i$&1300$^{i,k}$\\
CO$_2$&2300$^i$&2300$^{i,k}$\\
HCO&1600$^l$&1600$^l$\\
H$_2$CO&3700$^m$&3200$^{m}$\\
CH$_3$O&3700$^l$&3700$^l$\\
CH$_3$OH&3700$^n$&3700$^{n}$\\
N&720&720$^{f}$\\
N$_2$&790$^o$&1140$^p$\\
\hline\hline
\end{tabular}
%\multicolumn{3}{c}{
\\
$^a$\cite{cazaux2004}, $^b$\cite{bergeron2008}, $^c$\cite{Al-Halabi2007}, \cite{amiaud2007},$^d$\cite{Pirronello1997}, $^e$\cite{amiaud2006} for low coverage $^f$~\cite{minissale2015}, $^g$\cite{dulieu2013}, $^h$\cite{speedy1996}, \cite{fraser2001} energy derived of 5800~K with pre-factor of 10$^{15}$ s$^{-1}$ here corrected to have pre-factor of 10$^{12}$ s$^{-1}$, $^i$\cite{noble2012b}, $^j$\cite{borget2001}, \cite{minissalethesis2014},  $^k$\cite{Karssemeijer2014},$^l$\cite{garrod2006}, $^m$\cite{noble2012}, $^n$\cite{collings2004}, $^o$\cite{fuchs2006}, $^p$~extrapolated from \cite{kimmel2001} 
%\end{tabular}
%\end{center}
\end{table}

In the following subsections, we describe the different processes occurring on dust surfaces or in the bulk. 

\subsubsection{Accretion}
Species from the gas phase can collide with dust grains. The rate for this process (cm$^{-3}$ s$^{-1}$), which depends on the density of the species, their velocity, and the cross section of  the dust, can be written as
\begin{equation}
R_{acc}(X) = n_X v_X  n_{\rm{dust}} \sigma  S, 
\end{equation}
where n$_X$ is the density of the species X (cm$^{-3}$) and v$_X$ is the thermal velocity of the species X, which can be written as v$_X = \sqrt{\frac{8*k_{bz}*T_{\rm{gas}}}{\pi*amu_X}}$ cm s$^{-1}$. $S$ is the sticking coefficient of the species with the dust. Here we consider $S=1$, meaning that every species that arrives on dust sticks.
 
\subsubsection{Evaporation}
The species present on the dust surface can return to the gas phase because they evaporate. This evaporation rate depends on the binding energy of the species with the surface. However, the binding energies are very different if the species is bound to a bare or icy surface, as shown in Table~\ref{bind}. In our model, we define the fraction of the dust that remains bare as f$_{bare}$ and the fraction of the dust that is covered by ices as f$_{ice}$=1-f$_{bare}$  to apply the changes in binding energies as the dust becomes covered by ices. The evaporation rate (cm$^{-3}$ s$^{-1}$) of the species X can be written 
\begin{equation}
 R_{evap}(X)=n_X \nu \left(f_{bare}\exp(\frac{-E_{bare}(X)}{T_{\rm{dust}}})+f_{ice}\exp(\frac{-E_{ice}(X)}{T_{\rm{dust}}})\right),
\end{equation}
where $\nu$ is the oscillation factor  of the atom on the surface, which is typically  $\nu$=10$^{12}$ s$^{-1}$, and E$_{bare}$(X) and E$_{ice}$(X) the binding energies of the species X on bare and icy dust, respectively. When one monolayer is reached (4.4$\times 10^{-6}$ n$_H$ cm$^{-3}$), f$_{ice}$ becomes 1 and the binding energies and mobilities depend only on E$_{ice}$.

\subsubsection{Mobility} 
Surface species can move on grain surfaces through thermal hopping. The diffusion rate (in s$^{-1}$)  for a species X to move on bare surface or in the bulk can be written as
\begin{equation}
\alpha_{bare}({X}) = \nu  \exp({-\frac{2 E_{bare}(X)}{3 T}}) \\
\end{equation}
\begin{equation}
\alpha_{ice}({X}) = \nu \exp({-\frac{2 E_{ice}(X)}{3 T}}).
\end{equation}
We consider that the barrier for thermal hopping is two thirds of the binding energy. Moreover, as the dust surface evolves from bare to icy, we change the species mobility according to the physical-chemical conditions of surface (from bare to icy). In our model, we only consider mobility through thermal hopping. This is a good approximation for the mobility of H atoms at T$_{\rm{dust}}$$\ge$12K. Below this temperature, or if reactions involving chemisorption are considered, diffusion through tunneling should be taken into account (\citealt{cazaux2004}). We also do not consider the tunneling of O atoms in the bulk, which can be efficient until 25~K (\citealt{minissale2013}). 

%We should mention that we do not take into account microscopic reversability in this study (\citealt{cuppen2013}). The barriers to move from one place to another should be identical to the reverse barrier. In this work, we consider that species in the bulk have a binding energies independant of their positions. However, the binding energy should be the addition of the different binding energies between a species with its neighbours. Our rate equations model does not allow such subtilities and we will consider the detailed sites energies in a forthcoming work using Monte Carlo simulations. In this present study, we consider 2 possible surfaces, bare or icy, and assume that the binding energies for species moving from one site to another one are similar. This assumption has implications for the morphology, structure and composition of the ices. However, our purpose is to quantitatively estimate  the abundances of gas phase species and ices due to gas and grain surface chemistry. 

\subsubsection{Surface reactions}
Species on bare surfaces or in the bulk can meet and react to form products that will either stay on the surface or in the bulk, or be released into the gas phase. The different reactions and their associated barriers are reported as E$_a$ in Table~\ref{reac}. When the barrier for a reaction was not studied on the surface, we assumed that the reactions are similar to gas-phase reactions. %and used the NIST database (\citealt{manion2008}). 
The probability of chemical desorption is taken from Eq. 2 for bare grains (with values shown in Table~\ref{reac}). For icy surfaces, we considered the chemical desorption mechanism to be (1) reduced by one order of magnitude (with the exception of O+H, OH+H and N+N for which measurements were obtained) and (2) equal to zero.

The reactions occurring on dust surfaces are reported in Table~\ref{reac}. The products can be either released into the gas phase with a probability $CD_{bare}$ for reactions on bare surfaces, or $CD_{ice}$ for reactions on icy surfaces. The reaction rate (in cm$^{-3}$ s$^{-1}$) for the formation of the species X through encountering of species Y and Z can be written as
\begin{eqnarray}
 R_{\rm{gas}}(X)&=&\frac{n_Y  n_Z P_{reac}}{n_{\rm{dust}} n_{sites}} [f_{bare} CD_{bare}  (\alpha_{bare}(Y) + \alpha_{bare}(Z)) \nonumber \\
& & + f_{ice} CD_{ice} (\alpha_{ice}(Y) + \alpha_{ice}(Z))]
.\end{eqnarray}
If there is a barrier for the reaction to occur with an energy E$_a$ (in K), this barrier can be crossed by tunneling with a probability P$_{tun}$=$\exp(-2a*\sqrt{\frac{2\times m_{red} \times k_{bz} \times E_a}{\hbar^2}})$ or through thermal hopping with a probability P$_{therm}$=$\exp(-\frac{E_a}{k T_{dust}})$. In these expressions, a is the width of the barrier of 1 \AA\, and m$_{red}$ is the reduced mass of the reaction m$_{red}$=$\frac{m_Y\times m_Z}{m_Y + m_Z}$. P$_{tun}$ can be simply written as P$_{tun}$=$\exp(-0.406\sqrt{m_{red} \times E_a})$, with m$_{red}$ in atomic numbers and E$_a$ in kelvin. The probability of reaction, P$_{reac}$, depends on the rate for the species to react compared to the rate for the species to escape the site where they meet. This probability can be written as P$_{reac}$=$\frac{\nu (P_{tun}+P_{therm})}{\nu (P_{tun}+P_{therm})+ \alpha}$, where $\alpha$ is the mobility of the species X and Y. The rate R$_{\rm{gas}}$ directly gives the formation rate of the gas-phase species X in cm$^{-3}$s$^{-1}$ and can therefore be directly coupled with gas-phase reactions.

A reaction occurring on the dust surface can lead to products that remain on the surface. The probability of products remaining on the surface/bulk is noted as (1-$\mu$). The reaction rate (in cm$^{-3} s^{-1}$) for the formation of species X through encountering of species Y and Z can be written as
\begin{eqnarray}
 R_{surf}(X)&=&\frac{n_Y  n_Z P_{reac}}{n_{\rm{dust}}\times n_{sites}} [ f_{bare}(1-CD_{bare})  (\alpha_{bare}(Y) + \alpha_{bare}(Z)) \nonumber \\
 && +f_{ice}(1-CD_{ice}) (\alpha_{ice}(Y) + \alpha_{ice}(Z))]
.\end{eqnarray}

One of the reactions O+ H$_2$ $\rightarrow$ OH + H is energetically not accessible at low temperatures since the reaction has a barrier of 0.57 eV and is endoergic with an energy of 0.1 eV for H$_2$(v = 0) (\citealt{weck2006}). However, this reaction becomes exoergic for $v$ $>$ 0 and the barrier decreases to $\sim$0.4 eV for $v$ = 1 (\citealt{sultanov2005}). The reaction O($^3$P) + H$_2$ ($v$=1) proceeds through tunneling at low temperatures, as shown by \cite{weck2006}. In our model, we considered only the reaction of H$_2$ vibrationally excited with O($^3$P). For a population in equilibrium at a temperature T$_{\rm{dust}}$, the fraction of molecules in the J = 1 state is $\sim \exp({-\frac{150}{T_{\rm{dust}}}})$ $\sim 3 \times 10^{-6} $-- $6.7 \times 10^{-3}$ in the range of T$_{\rm{dust}}$ considered in this study (12 -- 30~K). 

%2 E_{bare}(X)}{3 T}})
%$\exp{−\frac{150}{T_{\rm{dust}}}}$
% $\sim$ 3 $\times$10$^{-6}$ -- 6.7 $\times$10$^{-3}$ in the range of T$_{\rm{dust}}$ considered in this study (12 -- 30~K). 

\subsubsection{UV photons and cosmic-ray-induced UV photons} 
Interstellar dust grains can be present in environments that are subject to radiation. Stars in the neighborhood can emit far-UV photons that impinge on the interstellar dust grains. When a UV photon arrives on an adsorbed species, it can be photodissociated. The products of the dissociation can be released directly into the gas phase, or one or both fragments can be trapped on the surface. \cite{andersson2008} calculated the outcomes of the photodissociation of water molecules in water ices. Once a water molecule is broken into OH and H in the first few monolayers of ice, several processes can occur with different probabilities if the ices are crystalline or amorphous (\citealt{andersson2006}). In most cases, H atoms are released into the gas phase and OH molecules are trapped (amorphous: 92~$\%$, crystalline: 70~$\%$) or H and OH are trapped (amorphous: 5~$\%$, crystalline: 14~$\%$). The most probable process is the trapping of OH. One of the possible processes is the reformation of water and its subsequent release into the gas phase. This process is called photodesorption (amorphous: 0.7~$\%$, crystalline: 1~$\%$). In our model, we considered that species being photodissociated by UV photons/cosmic ray-induced photons are trapped on the surface. Since we take into account the fact that species such as water can reform and be ejected into the gas through chemical desorption, the photodesorption of molecules is already considered in our model for water. For CO molecules, we considered CO photodesorption with a yield of 10$^{-2}$, which was obtained experimentally (\citealt{fayolle2011}). Furthermore, we considered that only the two first layers of ice can interact with photons or cosmic-ray-induced photons.

We considered that only the photons arriving directly on the species can photodissociate them, meaning that the cross section of reaction is similar to the one in the gas phase. We therefore used the values from the KIDA database to compute the dissociation rates of solid-phase species through photons and cosmic rays. We considered that dissociated solid species remain on the dust.

\subsection{Gas-phase reactions}
The numerical code Nahoon is publicly available on KIDA to compute the gas-phase chemistry for astrophysical objects (http://kida.obs.u-bordeaux1.fr/models/). A more detailed description of this model and the chemical network can be found in \cite{wakelam2012}. The grain surface chemistry presented in Sect. 2 was added to the existing gas-phase network, and we computed abundances of species in the gas phase and on grain surfaces as a function of time. \rm{The abundances derived in our simulations refer to fractional abundances with respect to the density of hydrogen n$_H$}\rm. We take into account pure gas-phase chemistry (bimolecular reactions, dissociative and radiative recombinations and associations, electron attachments, dissociations and ionizations induced by cosmic rays and UV photons), interaction between the species in the gas phase and the grain surfaces (adsorption, thermal desorption, desorption by cosmic rays, and chemical desorption), and grain surface reactions (chemical reactions and dissociation by UV photons and cosmic-ray-induced UV photons).

\subsection{Simulations}
In this section we perform several calculations to study the formation and abundances of O$_2$, HO$_2$ and H$_2$O$_2$ in astrophysical environments. Our study is motivated by the recent detections of O$_2$, HO$_2$ and H$_2$O$_2$ in $\rho$ Oph A (\citealt{bergman2011}; \citealt{parise2012}; \citealt{liseau2012}) and of O$_2$ in Orion (\citealt{goldsmith2011}). The detections of O$_2$ in these two distinct environments have been attributed to the effect of warm dust grains that allow the evaporation of the ices covering the dust. Additionally, the recent detection of H$_2$O$_2$ in $\rho$ Oph A, a molecule present in ices but never observed before, raised the question of its origin. \cite{du2012} modeled the abundances of gas and icy species in environments with similar physical characteristics as $\rho$ Oph A. These authors reproduced the abundance of H$_2$O$_2$ by assuming a chemical desorption efficiency of 7$\%$, derived from the approach of \cite{garrod2007}. Here we re-evaluate the abundances of O$_2$, HO$_2$ and H$_2$O$_2$ with the chemical desorption efficiencies derived in this study, and the change in these efficiencies when the dust surfaces are covered by ice. We performed three different simulations that are presented in the following subsections. Since the chemical desorption is very poorly defined on water ices (apart for the formation of OH, H$_2$O, and N$_2$), we performed an additional calculation without considering the chemical desorption on icy grains. Furthermore, to assess the impact of the chemical desorption on the chemical composition of the cloud, we performed a third calculation without the chemical desorption process. These simulations, called cloud 1, 2, and 3, are summarized in Table~\ref{cloud}.

\begin{table}
\caption{Summary of the simulations. \label{cloud}}
\begin{tabular}{|l|l|l|}
\hline
Simulation &  Chemical desorption & Chemical desorption \\
&  bare dust &  icy dust\\
\hline\hline
Cloud 1 & CD$_{bare}$=CD$_{bare}$(th) (eq. 2) & CD$_{ice}$=CD$_{bare}$(th)/10\\
&&CD$_{ice}$(OH)=0.25\\
&&CD$_{ice}$(H$_2$O)=0.3\\
&&CD$_{ice}$(N$_2$)=0.5\\
\hline
Cloud 2 &CD$_{bare}$=CD$_{bare}$(th) (eq. 2) &CD$_{ice}$=0\\
\hline
Cloud 3 &CD$_{bare}$=0 &CD$_{ice}$=0\\
\hline\hline
\end{tabular}
\end{table}

\subsubsection{Initial conditions}
\begin{figure}
\centering
\includegraphics[width=8.8cm]{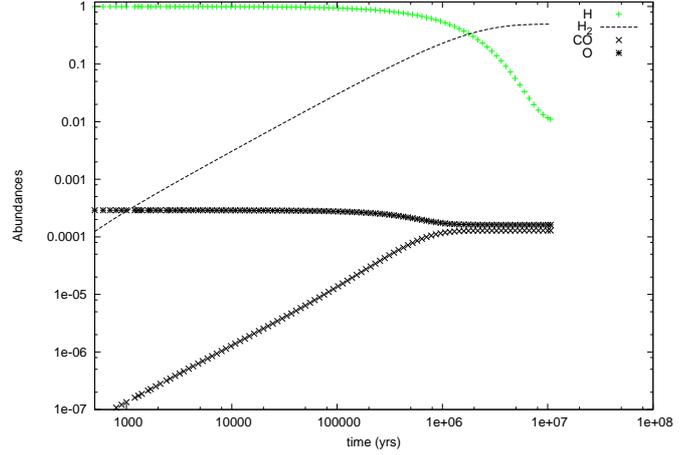}
\caption{Abundances of gas-phase species in our model cloud (A$_V$=3) as a function of time. \r,{The abundances obtained for long timescales provide the initial abundances for the cloud models at high extinctions.}\rm}
\label{gasinitial}
\end{figure}

The initial conditions for our cloud model are provided by simulating a cloud at an extinction of 3, for a density of n$_H$=10$^3$ cm$^{-3}$, G$_{0}$=10$^3$, T$_{gas}$=100~K, and T$_{dust}$=30~K. The model was run until 10$^7$ years, and the gas-phase abundances of the different species with time are reported in Fig.~\ref{gasinitial}. The chemical composition of the cloud reaches steady state at 10$^6$ years, and the abundances of atomic oxygen and CO are identical, in agreement with the abundances derived by \citealt{hollenbach2009}. 

\subsubsection{Cloud 1: with chemical desorption}
\begin{figure}
\centering
\includegraphics[width=8.8cm]{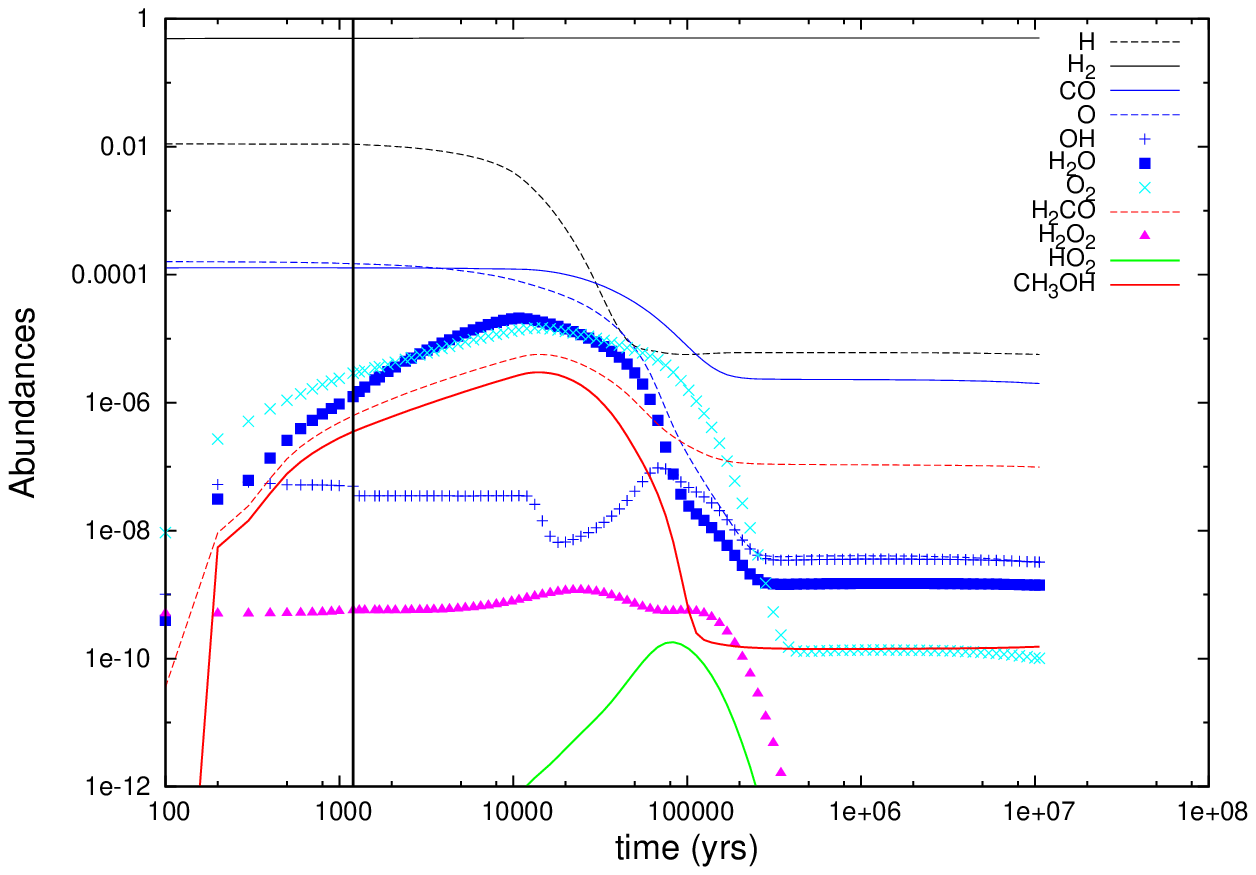}
\includegraphics[width=8.8cm]{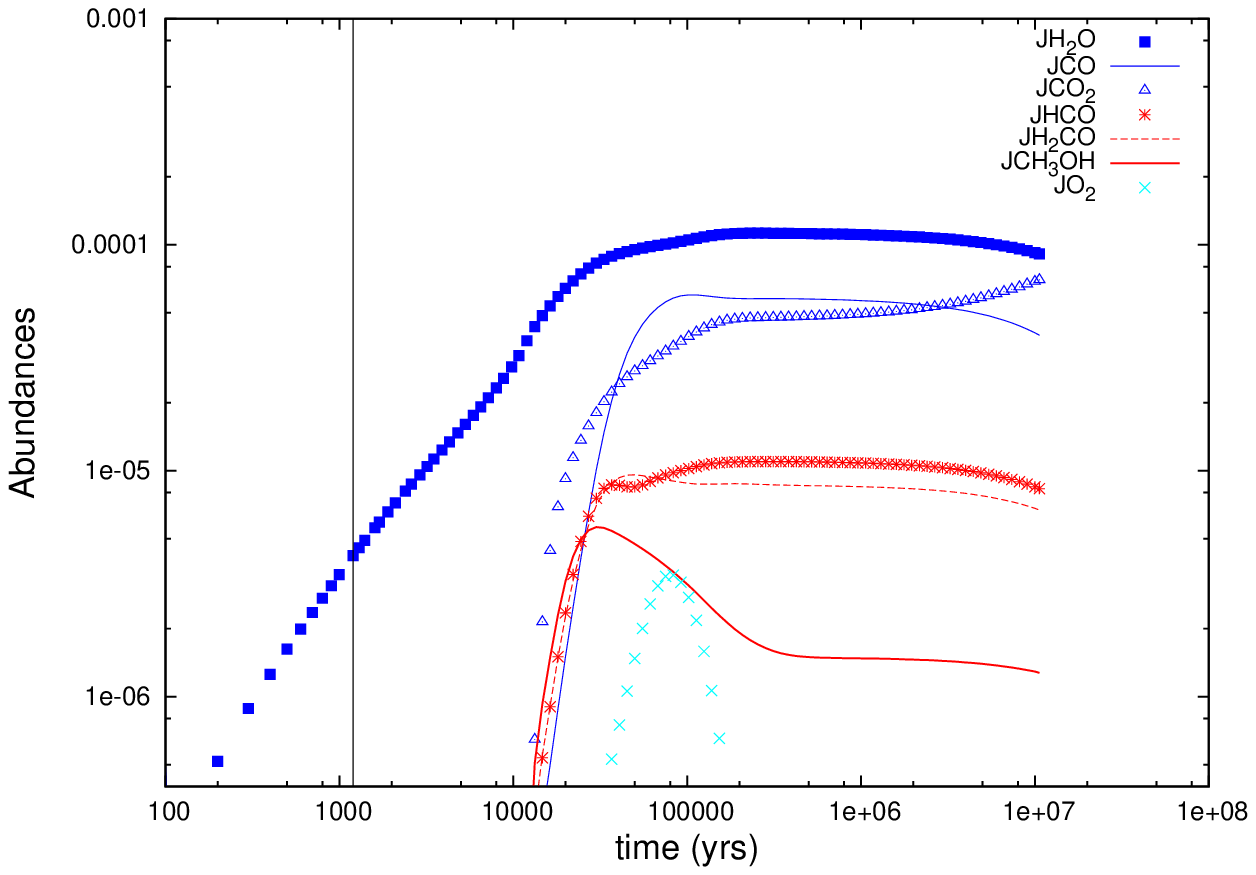}
\caption{Abundances of species in our model cloud 1 at A$_V$=15. The vertical line shows the transition where dust becomes covered with one layer of water ice. Top panel: Abundances of gas-phase species. Bottom panel: Abundances of species in the ices.}
\label{abun1}
\end{figure}

\begin{figure}
\centering
\includegraphics[width=8.8cm]{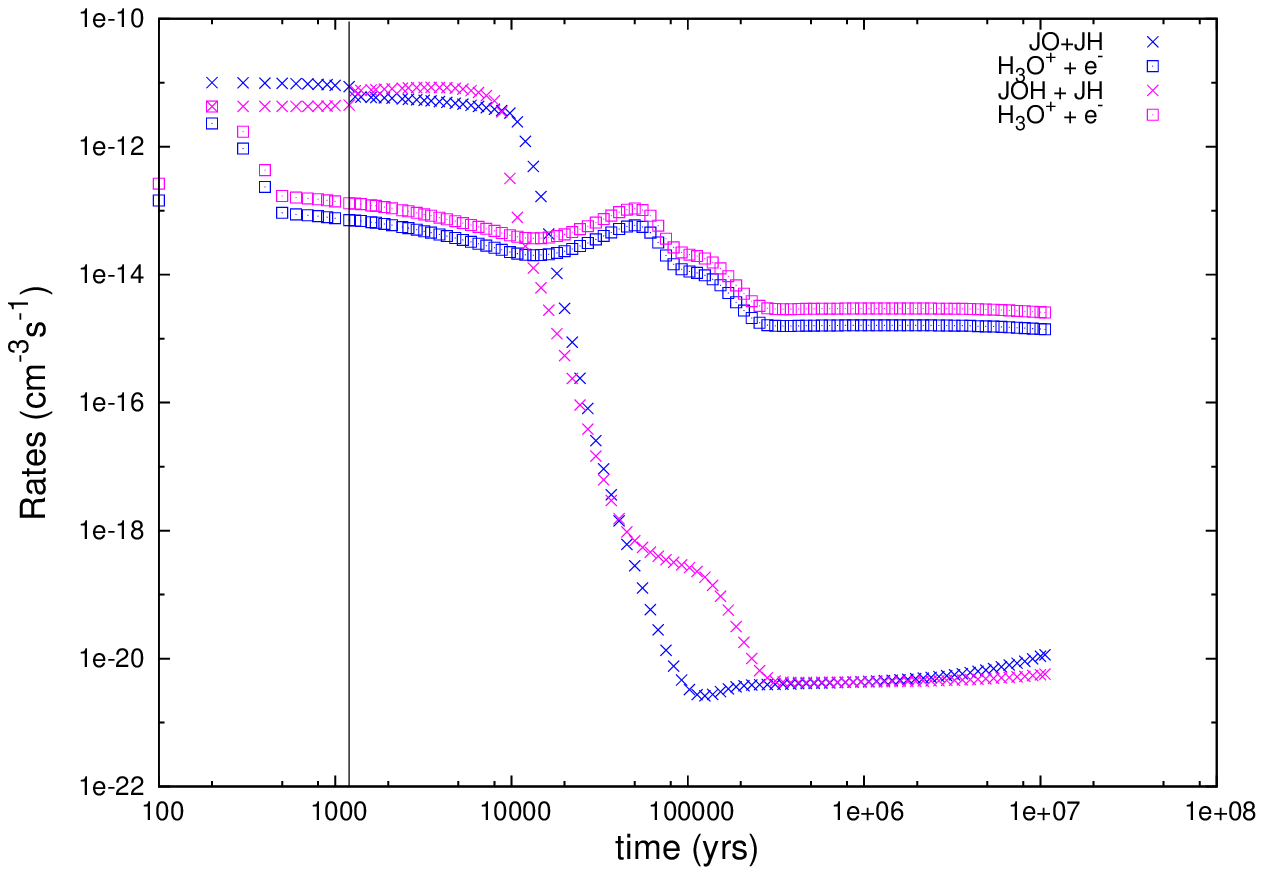}
\includegraphics[width=8.8cm]{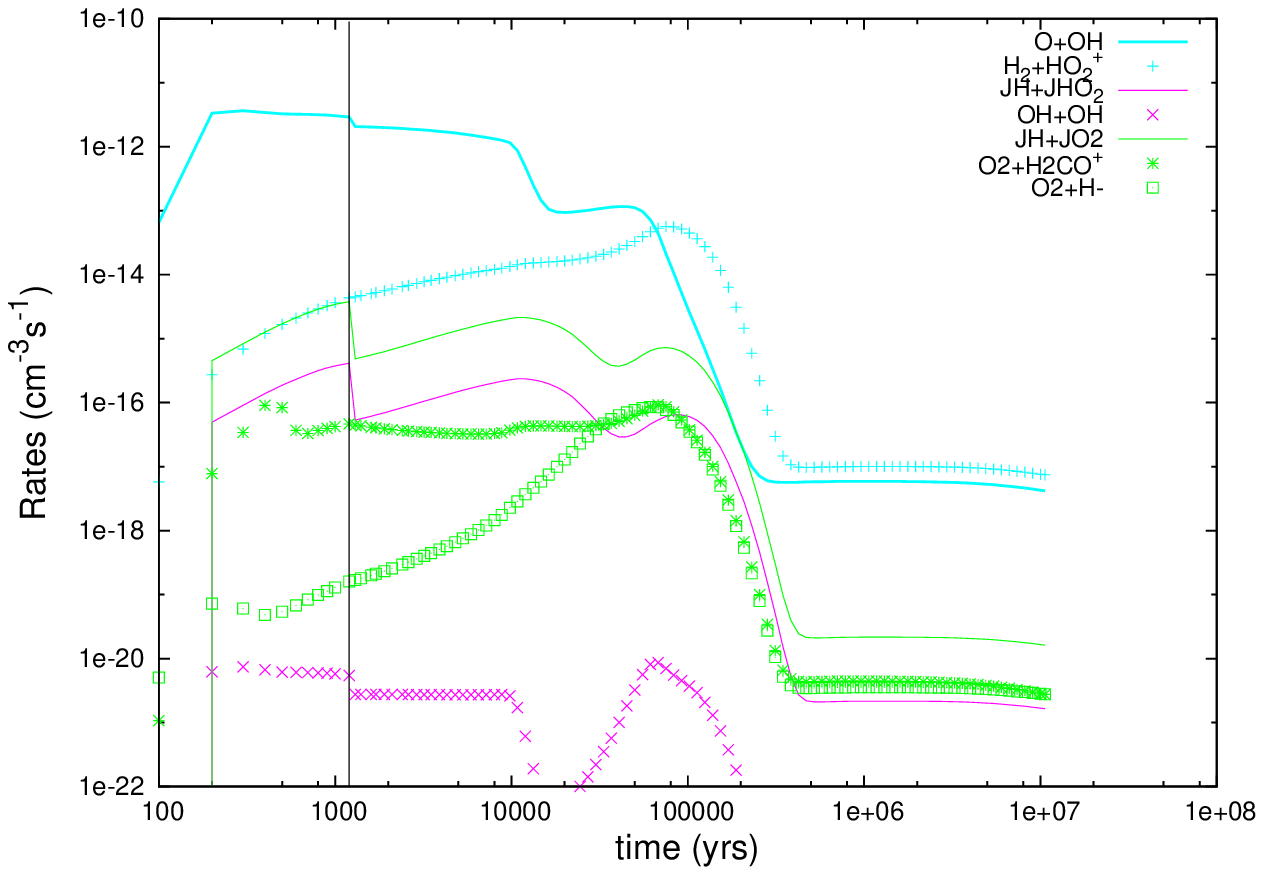}
\caption{Top: Formation rates of OH (blue) and H$_2$O (pink) through different routes. Bottom: Formation rates of O$_2$ (turquoise), H$_2$O$_2$(pink), and HO$_2$ (green) through the most dominant routes.}
\label{rate1}
\end{figure}

To follow the evolution of a molecular cloud, we used the abundances obtained with our calculations at Av=3 (for long timescale) as initial conditions. To mimic the physical conditions met in $\rho$ Oph A, we assumed a density of 10$^5$ cm$^{-3}$, a radiation field of G$_0$=10$^3$, an extinction of A$_V$=15, and a gas and dust temperature of  T$_{gas}$=T$_{dust}$=20~K, as assumed by \cite{du2012} taken from the observations from \cite{bergman2011}. We considered a chemical desorption efficiency as reported in Eq. 2 (and shown in Table~\ref{reac}) for bare dust, and ten times lower for icy dust apart for the formation of OH, H$_2$O, and N$_2$, which were derived experimentally, and were chosen as 0.25, 0,3 and 0.5, respectively (see Table~\ref{cloud} for a summary). The abundances in the gas phase and in the ices derived with our model are reported in Fig.~\ref{abun1}. \rm{To distinguish species in the gas phase and in solid phase (on dust or in the ices), we add the letter J in front of the name of the species when they are in solid phase (bottom panel of Fig.~\ref{abun1} where water is mentioned as JH$_2$O). }\rm The top panel of Fig.~\ref{abun1} shows the most abundant species in the gas phase as a function of time. The vertical line shows the time at which dust grains are covered by one layer of ice and therefore indicates the transition between submonolayer and multilayer regime \rm{(transition between bare and icy surfaces)}\rm. The abundance of O$_2$ increases with time until a maximum of 10$^{-5}$ is reached around 5$\times$10$^4$ years. OH in the gas phase is rapidly transformed into other species such as water, and therefore remains around 10$^{-7}$ until 10$^5$ years. Water increases up to few times 10$^{-5}$ in the early stages of the cloud evolution, until atomic hydrogen begins to be lacking in the cloud. Species such as H$_2$O$_2$ reach an abundance of $\sim$10$^{-9}$ until 10$^5$ years. HO$_2$, on the other hand, reaches a maximum of $\sim$ 10$^{-10}$ around 10$^5$ years. Methanol and H$_2$CO both increase with time until abundances of  6$\times$10$^{-6}$ and 2 $\times$10$^{-6}$ are reached at a few 2$\times$10$^4$ years. The coverage of species on dust is reported in the bottom panel of Fig.~\ref{abun1}. Water is accreted slowly with time and reaches one monolayer after $\sim$1200 years. This change from bare  to icy dust, shown as a vertical line, implies that species from the gas phase accreting onto dust are more closely bound to the surface (change in binding energies, as shown in Table~\ref{bind}). The chemical desorption efficiency is also strongly reduced, which implies that most of the species reacting on the surface (except OH, H$_2$O and N$_2$) desorb less upon formation. After water, solid CO$_2$ and CO become the most abundant species in the ice. This is due to the high temperature of the dust of 20~K. In this case, hydrogenation is less important than oxygenation since H atoms evaporate efficiently from the dust in comparison with oxygen atoms. The abundance of methanol in the ices is of about 1$\%$. This is due to the warm temperature at which oxygenation is favored compared to hydrogenation. 

The rates for the formation of OH (blue) and H$_2$O (pink) are reported in Fig.~\ref{rate1}, top panel. During the time that dust grains remain uncovered by ices (at $\sim$ 1200 years) and atomic H atoms are still abundant in the gas phase, formation of OH and H$_2$O in the gas phase is dominated by reaction on surfaces and subsequent release in the gas through the chemical desorption process. The reaction of adsorbed hydrogen with adsorbed oxygen and adsorbed OH enhances the gas-phase OH and H$_2$O. This gas enrichment through solid-state reactions dominates until $\sim$10$^4$ years. After this time, the formation of OH and water is dominated by recombination of H$_3$O$^+$ , which can either form OH + H$_2$ or H$_2$O + H.

The rates for the formation of O$_2$, H$_2$O$_2$ and HO$_2$ are reported in the bottom panel of Fig.~\ref{rate1}. The formation of O$_2$ (turquoise) occurs mostly through gas-phase processes that involve OH. In this sense, O$_2$  is a direct by-product of the enhanced formation of OH through the chemical desorption process on dust. The formation H$_2$O$_2$ (pink), on the other hand, is dominated by dust reactions of adsorbed hydrogen with adsorbed HO$_2$ leading to the formation and desorption of H$_2$O$_2$. This process is very efficient on bare dust, until $\sim$ 1200 years, and decreases by one order of magnitude once the dust becomes covered by ices. Even with a much lower efficiency, the chemical desorption process dominates the gas-phase ion-neutral reactions. Therefore, the H$_2$O$_2$ molecules present in the gas phase are due to the chemical desorption process, which dominates in the early time of the cloud evolution, when dust grains are still bare. The formation of HO$_2$ (green) is dominated by the hydrogenation of O$_2$ on dust and its subsequent release into the gas through chemical desorption. After dust becomes covered by ice, the chemical desorption process becomes much less efficient, but still strongly dominates the ions-neutral reactions. This reduced efficiency still allows an increase in gas-phase HO$_2$ until abundances of 10$^{-10}$ are reached.

\subsubsection{Cloud 2: without chemical desorption on icy dust}

\begin{figure}
\centering
\includegraphics[width=8.8cm]{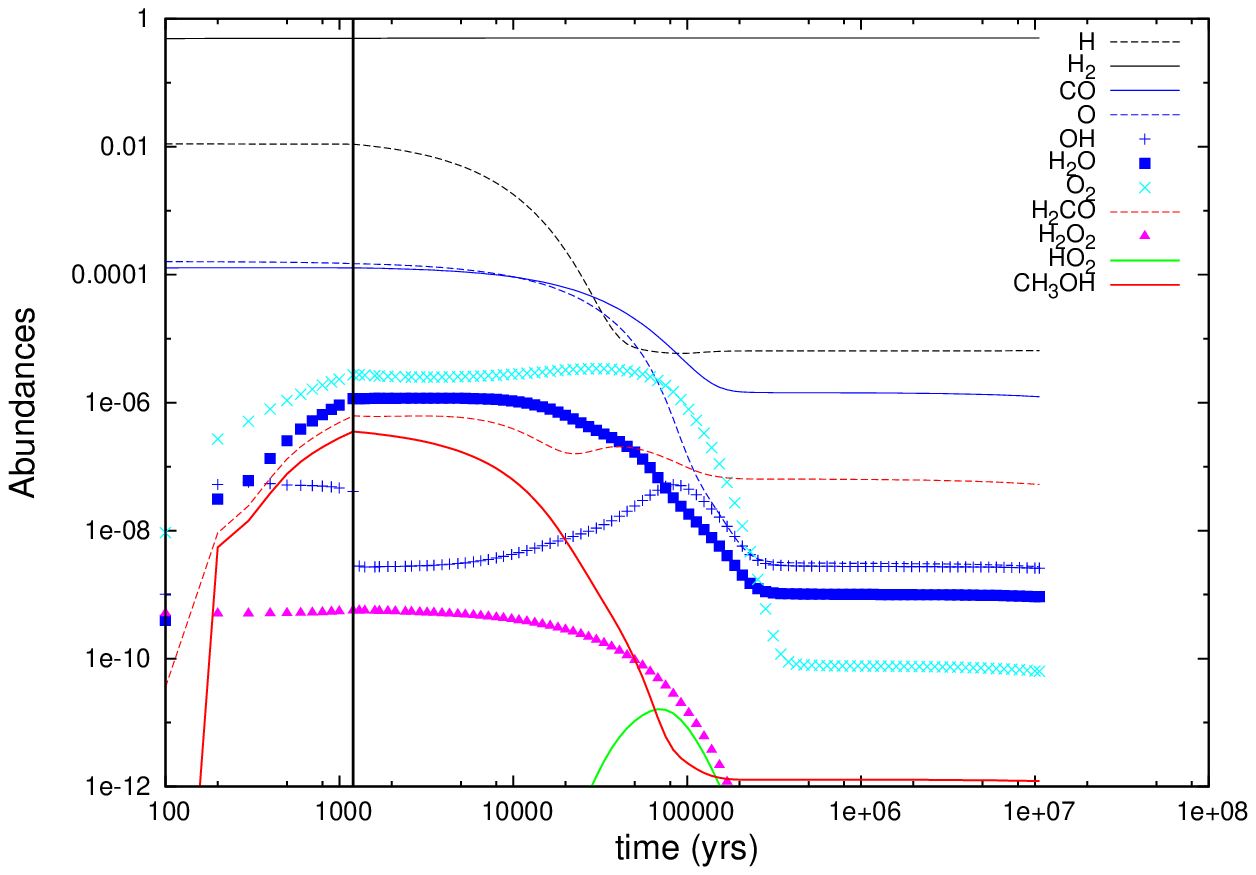}
\includegraphics[width=8.8cm]{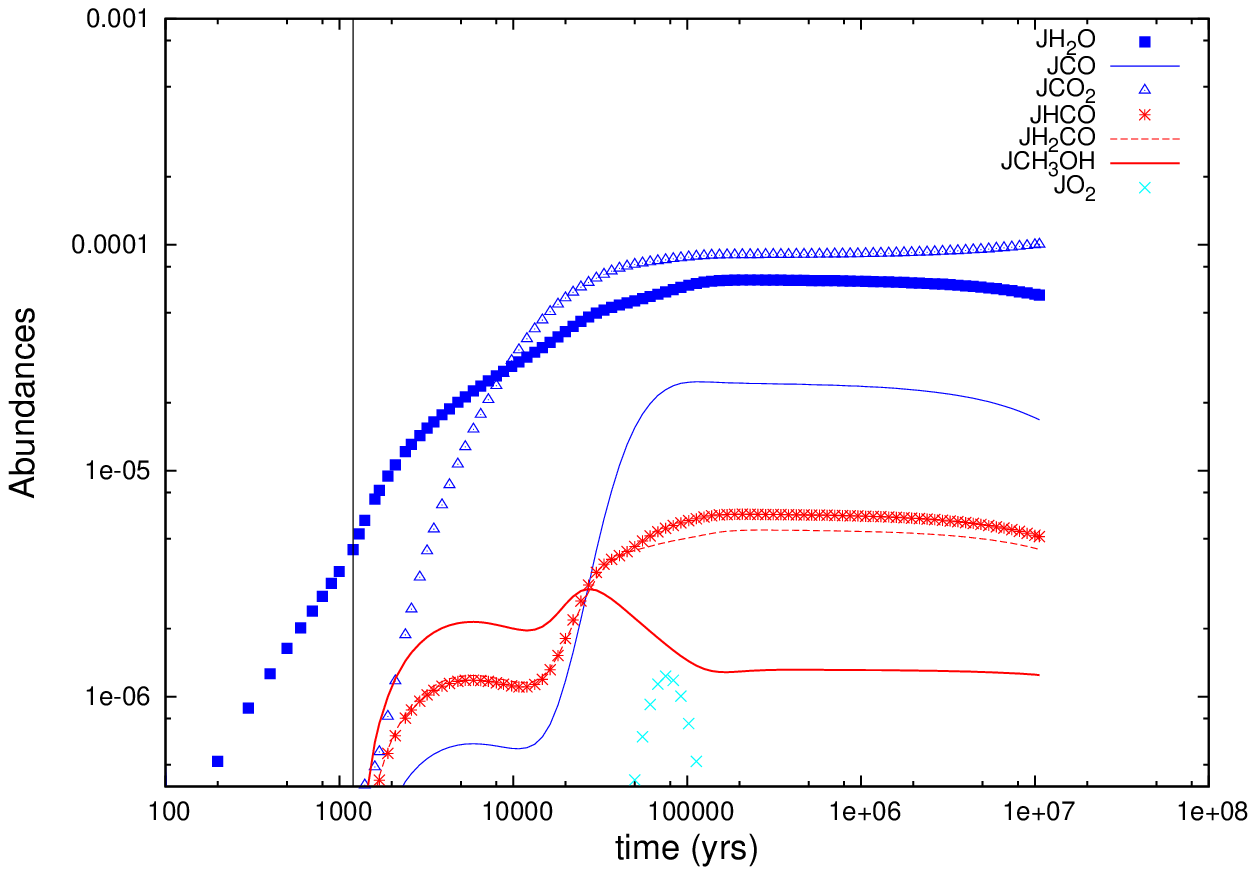}
\caption{Abundances of species in our model cloud 2 at A$_V$=15. The vertical line shows the transition where dust becomes covered with one layer of ice. Top panel: Abundances of gas-phase species. Bottom panel: Abundances of species in the ices.}
\label{cdi0}
\end{figure}

In our second simulation, we considered the chemical desorption on icy dust to be negligible (we set CD$_{ice}$=0, see Table~\ref{cloud}). These simulations allow determining the lower limits of the impact of the chemical desorption process because chemical desorption on icy dust is not well defined. The abundances of the species in the gas phase are reported in the top panel of Fig.~\ref{cdi0}. While the abundances of oxygen, hydrogen, and CO are similar to our previous calculations, many species are formed in the early stages of cloud evolution (in the submonolayer regime, before 1200 years, as presented by the black vertical line), and only decline at later stages. This is the case for H$_2$O$_2$, H$_2$CO, and methanol, which are formed through dust grain reactions and are released into the gas phase through chemical desorption as dust grains are still bare. Once dust becomes covered by ice mantles (time $>$1200 years), many species are not produced efficiently, such as OH, which declines abruptly and in turn diminishes the abundance of O$_2$. HO$_2$, which was mostly formed though chemical desorption from icy grains in model cloud 1, has an abundance that decreases by almost one order of magnitude. The coverage of species on dust is reported in the bottom panel of Fig.~\ref{cdi0}. Water is the main component of the ices until 10$^4$ years, after which CO$_2$ becomes more abundant. The abundances of solid CO, O$_2$, H$_2$CO, and methanol decrease compared to the abundances reported for model cloud 1. The chemical desorption process not only affects the composition of the gas phase, but also the composition of the solid state. The chemical desorption for the formation of OH and water on icy dust allows an enrichment of the gas phase in these species. Therefore, water can accrete onto the dust, which increases the abundance of water in solid form. By neglecting the chemical desorption on icy dust, the ices are lacking in water ices, and CO$_2$ becomes a dominant species.

\subsubsection{Cloud 3:  without chemical desorption}
\begin{figure}
\centering
\includegraphics[width=8.8cm]{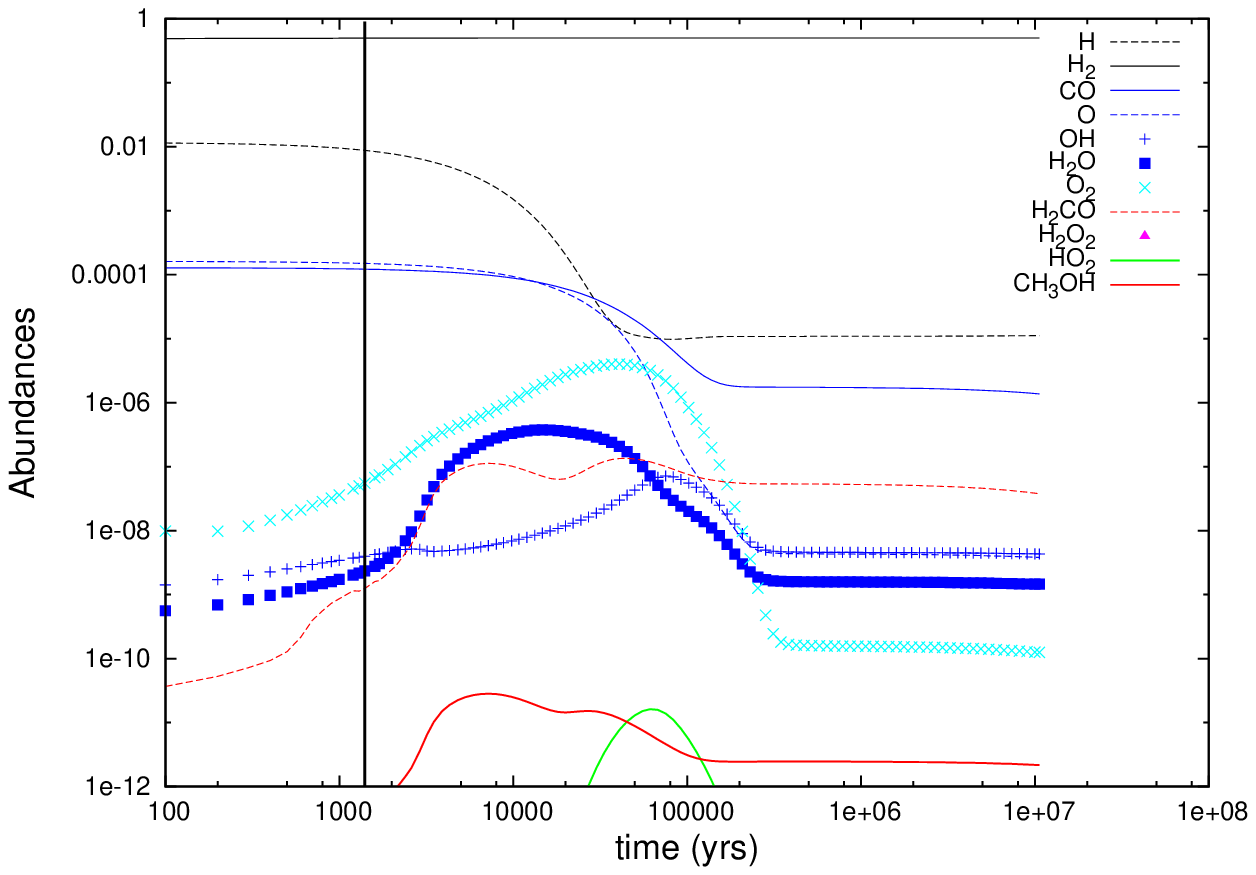}
\includegraphics[width=8.8cm]{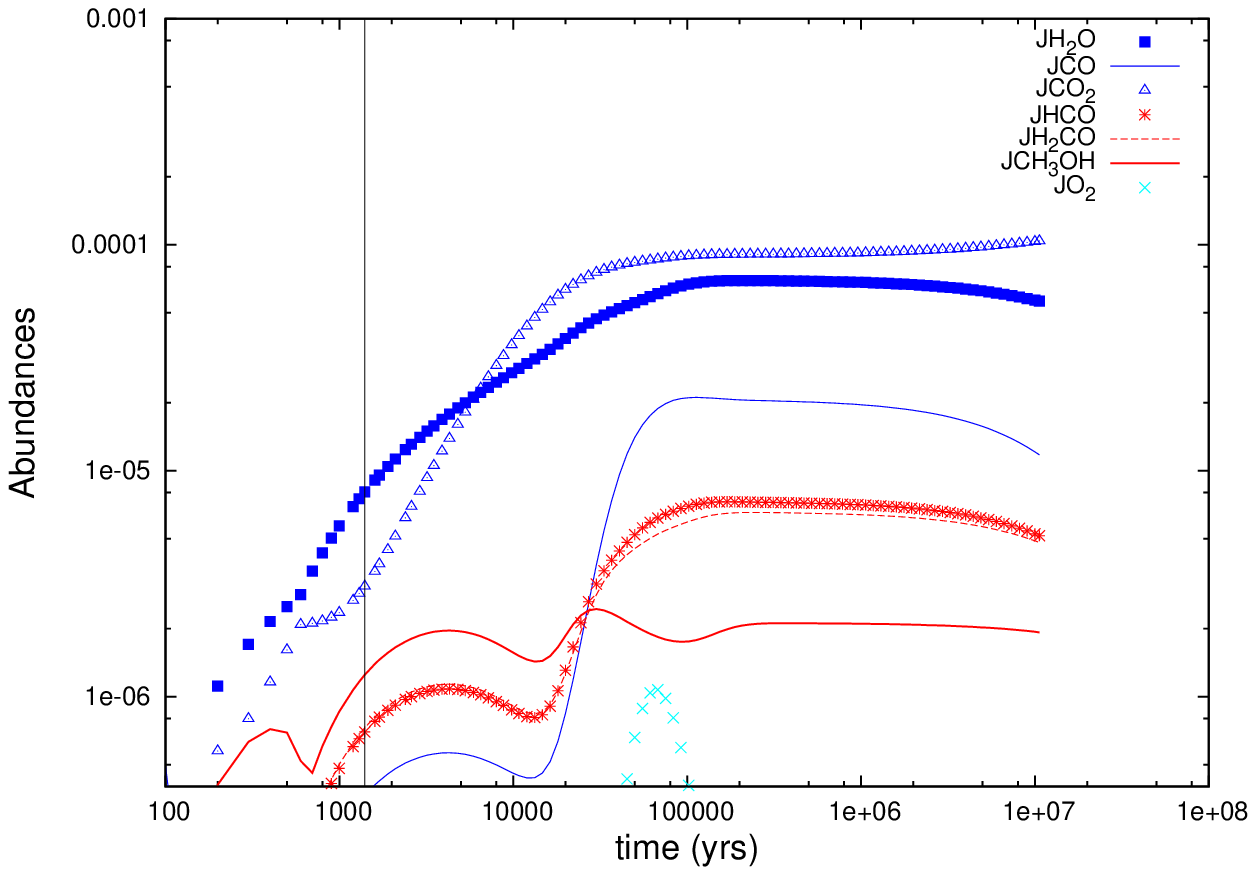}
\caption{Same as Fig.~\ref{abun1} without the chemical desorption process. Abundances of species in our model cloud 3 at A$_V$=15. The vertical line shows the transition where dust becomes covered with one layer of ice. Top panel: Abundances of gas-phase species. Bottom panel: Abundances of species in the ices.}
\label{cd0}
\end{figure}

To assess the impact of the chemical desorption process on the chemical composition of our cloud model, we performed an additional simulation without chemical desorption. The abundances of the species in the gas phase are reported in the top panel of Fig.~\ref{cd0}.  While the abundances of oxygen, hydrogen, and CO are similar to the one derived in our previous calculations, the abundance of methanol is decreased by four orders of magnitude, while H$_2$O$_2$ is absent from the figure since its abundance is $\sim$ 10$^{-14}$. Water and OH abundances are also decreased in the early stages of the cloud evolution, since their formation is not ensured by the chemical desorption process, but by less efficient gas-phase routes as shown in Fig.~\ref{rate1}. Since O$_2$ formation involves OH molecules, the lower OH abundance diminishes the abundance of O$_2$, as in the case of model cloud 2. The coverage of species on dust is reported in the bottom panel of Fig.~\ref{cd0}. Water molecules that form on the dust do stay on the dust. The building-up of the ices is somewhat faster ($\sim$1000 years), while the abundances of water in the gas is lower than in our previous simulations. Water ice dominates the solid state in the early stages of the cloud evolution, whereas CO$_2$ becomes the main constituent of the solid phase in the latter stage after 10$^4$ years. The other solid species behave in a similar manner as in the case of model cloud 2. To summarize, our simulations show that the chemical desorption process is needed to account for the presence of some species into the gas phase, such as H$_2$O$_2$ and methanol. The chemical desorption process also has an important impact on the chemical composition of interstellar ices.

\begin{figure}
\centering
\includegraphics[width=8.8cm]{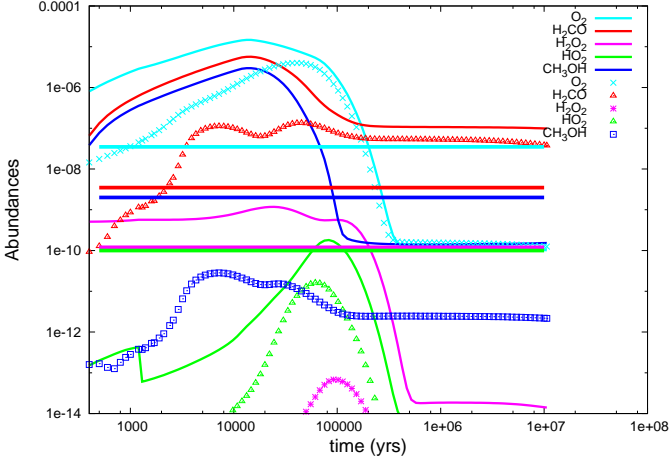}
\caption{Abundances of O$_2$ (cyan), H$_2$CO (red), H$_2$O$_2$ (pink), CH$_3$OH (blue), and HO$_2$ (green) in our model clouds taking into account the chemical desorption process (model cloud 1: lines) and without the chemical desorptions process (model cloud 3: dots). The horizontal lines show the observations of $\rho$ oph A.}
\label{obs}
\end{figure}

%\begin{figure}
%\centering
%\includegraphics[width=8.8cm]{gascom.eps}
%\caption{Abundances of O$_2$ (cyan), H$_2$CO (red), H$_2$O$_2$ (pink), CH$_3$OH (blue) and HO$_2$ (green) in our model clouds taking into account the chemical desorption process (model cloud 1: lines) and without taking into account the chemical desorptions process (model cloud 3: points). The horizontal lines show the observations of $\rho$ oph A.}
%\label{obs}
%\end{figure}

\section{Discussion}
We showed that for the reactions considered in this study, the chemical desorption process is well constrained on bare dust, but is difficult to quantify on icy surfaces since its efficiency is strongly reduced. To account for the uncertainties coming from the ill-defined process on icy surfaces, we performed simulations considering the chemical desorption efficiency on icy dust to be either very low (tenth of its value on bare dust) or zero, or the efficiency to be zero on bare and icy dust. In Fig. \ref{obs} we report the abundances of the molecules that have been observed in $\rho$ Oph A. We show the abundances with (lines) and without
(dots) chemical desorption. Our simulations show that the presence of species such as H$_2$O$_2$ and methanol can be enhanced by four orders of magnitude due to the chemical desorption process, while this enhancement is one order of magnitude for species such as HO$_2$ and H$_2$CO. This enhancement is very important in the earliest stages of cloud evolution (until 10$^4$ years), when dust grains are still bare and the chemical desorption efficiency is very efficient. The species formed at this epoch can remain in the gas phase for very long timescales (until 10$^5$ years), until all the species freeze onto the dust grains. The presence of HO$_2$, H$_2$O$_2$ and methanol indicate an enhancement through solid-phase chemistry and the imminent freeze-out of the cloud.
%Therefore, the presence of species such as HO$_2$, H$_2$O$_2$ and methanol, are relics from the youth of the cloud and can be seen in the time preceeding the freezing of the cloud. 

Observations of H$_2$O$_2$ and HO$_2$ in $\rho$ Oph A show that the abundances of these species are $\sim$ 10$^{-10}$ each (\cite{bergman2011}; \cite{parise2012}). Methanol has also been observed with an abundance of $\sim$ 10$^{-9}$ \cite{liseau2003, bergman2011}. The observed abundances of these molecules are reported as solid horizontal lines in Fig.~\ref{obs}. The abundances of O$_2$, H$_2$O$_2$, HO$_2$ , and CH$_3$OH are well reproduced by our model cloud 1 (lines) between 10$^5$ and 3$\times$ 10$^5$ years, while our cloud model 3 (without chemical desorption) underestimates the abundances by one to three orders of magnitude for HO$_2$ and H$_2$O$_2$ , respectively. To reproduce the abundances observed in $\rho$ Oph A, the chemical desorption mechanism is therefore
required, as was shown by \citealt{du2012}. We here reconsidered the chemical desorption mechanism by using the efficiencies that we derived experimentally. We also considered the changes in binding energies as the dust becomes covered by icy mantles. In an environment with physical characteristics similar to $\rho$ Oph A, dust grains are warm enough ($\sim$ 20~K) to allow the formation of oxygenated species on dust surfaces, which leads to the formation of HO$_2$ and H$_2$O$_2$. We showed that the presence of H$_2$O$_2$ is due to the first phases of the evolution of a cloud, when dust grains are bare and chemical desorption is very efficient. For methanol, chemical desorption on bare dust and icy dust is required to account for the abundances that are observed in $\rho$ Oph A of $\sim$10$^{-9}$. We also showed that the abundances of HO$_2$ observed in the gas indicate an enrichment through chemical desorption throughout the evolution of the cloud. Our simulations showed that HO$_2$, H$_2$O$_2$ , and methanol can all be present in interstellar gas around 10$^5$ years in conditions similar to $\rho$ Oph A, as shown in Fig.~\ref{obs}. 

\section{Summary and conclusions}

In part I of this study, a collection of experiments has been reported to quantify the fraction of species forming on dust that is ejected into the gas phase upon formation, the so-called chemical desorption process. An analytical expression that we reported here was derived to quantify the efficiency of this process on different substrates such as oxidized graphite and amorphous silicates. This formula depends on the equipartition of the energy of newly formed products and the binding energy of the products, and it reproduces the experimental results on bare surfaces well. However, on icy surfaces the chemical desorption process is strongly reduced and cannot be quantified by our experimental results. 

To address the importance of the chemical desorption process on the composition of interstellar gas, we used a three-phase chemical model, which combines gas-phase chemistry and surface and bulk chemistry, and we used our analytical formula to account for the chemical desorption process. We computed the chemical composition of a cloud with similar physical characteristics as $\rho$ Oph A to reproduce species such as HO$_2$ and H$_2$O$_2$ , which have been observed in this environment ((\citealt{bergman2011}; \citealt{parise2012}). We show that the presence of HO$_2$ and H$_2$O$_2$ in interstellar gas indicates that species are starting to freeze out. H$_2$O$_2$ is produced through the chemical desorption mechanism in the earliest stages of the cloud evolution, when dust grains are still bare, while HO$_2$ is abundant only for a very short period of time, preceding the freeze-out. The presence of these species, enhanced by the chemical desorption process, is an indicator of the imminent freeze-out of the cloud.

\begin{acknowledgements}
S. C. is supported by the Netherlands Organization for Scientific Research (NWO; VIDI project 639.042.017) and by the European Research Concil (ERC; project PALs 320620). We acknowledge the support of the French National PCMI program funded by the CNRS and the support of the DIM ACAV, a funding program of the R\'egion Ile de France. We would like to thank the anonymous referee for the comments and suggestions that considerably improved our manuscripts (part I and part II).
\end{acknowledgements}

\end{document}